# STABILITY ANALYSIS AND μ-SYNTHESIS CONTROL OF BRAKE SYSTEMS

S. Lignon*, J-J. Sinou, and L. Jézéquel.

Laboratoire de Tribologie et Dynamique des Systèmes UMR CNRS 5513
Ecole Centrale de Lyon, 36 avenue Guy de Collongues, 69134 Ecully, France.

## ABSTRACT

The concept of friction-induced brake vibrations, commonly known as judder, is investigated. Judder vibration is based on the class of geometrically induced or kinematic constraint instability. After presenting the modal coupling mechanism and the associated dynamic model, a stability analysis as well as a sensitivity analysis have been conducted in order to identify physical parameters for a brake design avoiding friction-induced judder instability.
Next, in order to reduce the size of the instability regions in relation to possible system parameter combinations, robust stability via μ-synthesis is applied. By comparing the unstable regions between the initial and controlled brake system, some general indications emerge and it appears that robust stability via μ-synthesis has some effect on the instability of the brake system.

## NOMENCLATURE

| | |
|---|---|
| **C** | damping matrix |
| **G** | controller |
| **K** | stiffness matrix |
| **M** | mass matrix |
| $N$ | normal load |
| **P** | initial system |
| $T$ | tangential load |
| $x$ | scalar |
| **x** | vector |
| $\dot{\mathbf{x}}$ | vector of velocity |
| $\ddot{\mathbf{x}}$ | vector of acceleration |
| $\boldsymbol{\Delta}_{set}$ | structured uncertainties set |
| $\lambda$ | eigenvalue of the nominal system |
| $\tilde{\lambda}$ | eigenvalue of the controlled system |
| $\mu_f$ | brake friction coefficient |
| $\mu$ | structured singular value |

# 1  INTRODUCTION

Friction-induced vibration and instability are complicated phenomena that have been studied in detail by many researchers [1-23]. However, they are still a major concern in a wide range of mechanical systems due to the difficulty in resolving the problem. This is especially the case for brake systems, where friction-induced vibration due to coupling modes can cause severe damage or/and noise. So the prevention and prediction of unstable vibrations are actually very complex and important problems for the vehicle brake industry. In order to avoid these problems, the effects of some specific system parameters (typically mass, stiffness, damping,…) need to be studied in order to detect the stable and unstable zones of the mechanical system subject to friction-induced instability. Though many studies have been conducted and some of them have been successfully applied to particular brake systems and running conditions, it can be very difficult to find suitable values of the system parameters in order to obtain stable brake systems for all operating conditions. In these cases, the engineer therefore needs to find suitable devices to control instability in the brake system.

In recent decades, friction-induced vibration has received considerable attention from a number of researchers: Ibrahim [1-2], Bowden and Tabor [3], Rabinowitz [4], Armstrong–Hélouvry [5], and Oden and Martins [6]. Their investigations were conducted in order to find different mechanisms of friction-induced system instability. This type of analysis was then introduced in the context of brake noise to predict the dynamic behaviour of brake systems and to prevent instability (Ouyang *et al.* [12], North [7-8], Kinkaid *et al.* [10], etc).

In order to find the most suitable mechanism to describe friction-induced vibration in brake systems, these different mechanisms have to be examined. They fall into four classes: stick-slip, variable dynamic friction coefficient, sprag-slip [22] and geometric coupling of degrees of freedom [9, 15-21]. The sprag-slip action was described by Spurr [22] and does not depend on a friction coefficient varying with the relative rotation speed of the brake disc. Next, a number of investigations have been developed by considering kinematic constraint or geometric instability. This mechanism involves the coupling of the different degrees of freedom. It can be seen as an extension of the idea of the sprag-slip model [22]. Earles *et al.* [15, 16, 18, 20] and North [7-8] conducted extensive studies of kinematical constrained instability models. They demonstrated that instability may occur even if the friction coefficient is constant.

In this study, a modal coupling mechanism involving two system modes coupled together due to the friction interface will be considered. This instability may be defined as a geometrical coupling where two system modes move closer in frequency as the friction coefficient increases.

In this study, the application of robust control via µ-synthesis for a brake system is tested in order to avoid instability or in order to reduce the instability regions. In the first section, some basic concepts of µ-synthesis will be introduced. In the second section, the modal coupling mechanism used in this study will be briefly presented and the application of µ-synthesis for judder instability will be investigated. Next, a stability analysis and some interesting studies of possible system parameter combinations for the initial and controlled brake systems will be undertaken in order to examine the varying effects of robust control analysis on the size of the instability regions. This sensitivity analysis will be conducted in order to find the physical parameters for a brake design which avoid friction-induced instability in the case of controlled and uncontrolled brake systems. Finally, some natural extensions and possible applications of this methodology will be briefly described in the conclusions.

# 2 µ-SYNTHESIS

## 2.1 INTRODUCTION

The robustness of a system **P** with uncertainties represented by a set $\Delta_{set}$ of block-diagonal matrices is studied with the non singularity of the matrix $\mathbf{I} - \mathbf{P}\Delta$ (where **I** is the identity matrix), for $\Delta \in \Delta_{set}$. In order to treat this problem, the structured singular value $\mu$ is introduced; this parameter $\mu$ will be defined below.

This theory was introduced by Doyle [24] in 1982 and has become a standard tool in the robustness analysis of linear systems. It directly considers the problem of robust stability for a known plant subject to a block-diagonal structured uncertainty connected in feedback. The utility of $\mu$ lies in the fact that essentially any block diagram interconnection of systems and uncertainties may easily be rearranged into this standard form, i.e. where the uncertainty structure is block-diagonal.

## 2.2 STRUCTURED SINGULAR VALUE

This section defines the structured singular value $\mu(.)$. We consider matrices $\mathbf{P} \in \mathbf{R}^{n \times n}$ and introduce a structure $\Delta_{set}$ to define $\mu(\mathbf{P})$. This structure $\Delta_{set}$ is a prescribed set of block-diagonal matrices and may be defined differently for each problem depending on the uncertainty of the problem.

By definition, $\mu(\mathbf{P})$ is defined for $\mathbf{P} \in \mathbf{C}^{n \times n}$ by:

$$\mu(\mathbf{P}) = \frac{1}{\min\{\bar{\sigma}(\Delta) : \Delta \in \Delta_{set}, \ \det(\mathbf{I} - \mathbf{P}\Delta) = 0\}} \quad (1)$$

unless no $\Delta \in \Delta_{set}$ makes $\mathbf{I} - \mathbf{P}\Delta$ singular, in which case $\mu(\mathbf{P}) = 0$. $\bar{\sigma}(\Delta)$ corresponds to the maximum singular value of the matrix $\Delta$.

$\mu$ is then a function of two variables: the complex matrix **P** and the structure $\Delta_{set}$.

Considering the loop shown in Figure 1, $\mu(\mathbf{P})$ can be interpreted as a measure of the smallest uncertainty (represented by the matrix $\Delta$) that causes instability of the constant matrix feedback loop. The norm of this destabilizing $\Delta$ is exactly $\frac{1}{\mu(\mathbf{P})}$. It means that the weaker $\mu(\mathbf{P})$ is, the more robust the system.

Details concerning the calculation of the structured singular value are given in Packard and Doyle [25].

## 2.3 µ-SYNTHESIS

The definition of $\mu$ allows an extension of the stability analysis of systems by considering the system illustrated in Figure 2. This system is composed of three blocks $\Delta$, **P** and **G** that define the perturbation matrix, the initial system which should be controlled, and the controller, respectively.

The input / output couples are $(u_0, Y_0)$, $(u_1, Y_1)$ and $(u_2, Y_2)$ which define respectively the perturbation variables associated with the perturbation matrix $\Delta$, the measurable output (with the input control $u_1$) and the performance variables where $u_2$ includes the commands and the excitation and $Y_2$ represents the errors and the results.

Then, µ-synthesis consists of determining the controller **G** allowing the stability of the system in the presence of the uncertainty $\Delta$. The resolution is conducted by applying successive iterations. As the controller is often denoted by **K** (notation already used here for the stiffness matrix), this resolution is called the D-K iterations. These iterations are repeated unless $\mu < 1$. If $\mu < 1$, the robust stability of the system is assured for the given uncertainties. The theory of µ-synthesis is developed in Packard *et al.* [26], Venini [27], Balas *et al.* [28] and Markerink *et al.* [29]. Some applications can be found in Lanzon [30], Wu and Lin [31].

## 3    BRAKE SYSTEMS

In order to demonstrate the suitability of µ-synthesis to brake systems and in order to link the effect of specific parameter variation with stability of the design features, a parameter model including friction forces at the rubbing surface and mechanisms for friction-induced system instability is established and the equations of motion are determined.

The problem considered in this study deals with a modal coupling mechanism [11-12] that results from the coupling of two system modes due to the friction interface. The first mode corresponds to the suspension mode of the front axle assembly and the second mode corresponds to the normal mode of the brake piston elements. This phenomenological model was established through experimental investigations [32] and the friction-induced vibration was observed in the $50-100$ Hz range without variation of the brake friction coefficient. The fact that instability may occur even if the coefficient of friction is constant is a very common phenomenon that has been observed by many researchers [9,11, 15-21].

In the following sections, two analytical models (the initial and controlled systems) will first be presented.  Second, a stability analysis for each system will be undertaken and the initial and controlled systems will be compared in order to demonstrate the suitability of robust control for brake systems.

### 3.1    INITIAL SYSTEM

The initial system studied here is modelled as a three-degrees-of-freedom system, as illustrated in Figure 3(A): translational and normal displacement in the y-direction of the mass $m_2$ defined by $Y_2(t)$ and $X_2(t)$, respectively, and the translational displacement in the x-direction of the mass $m_1$ defined by $X_1(t)$. As previously explained, each mode is linked to a single vibration mode of the brake system: $(k_1, m_1)$ and $(k_2, m_2)$ define the dynamic of the brake piston elements and the dynamic of the suspension mode of the front axle assembly, respectively. The modal coupling mechanism involves the two modes coupled together due to the friction interface. This mechanism may induce a classic flutter instability where two solutions for the dynamic behaviour of the mechanical system exist. The first solution is an unstable equilibrium whereas the second is a periodic solution. Then, any perturbation of the equilibrium point implies self-excited vibrations.

In order to simulate the modal coupling mechanism due to the friction interface, this friction interface slopes with an angle $\theta$. This assumption may be seen as a geometric coupling with the braking system. This slope couples the normal and tangential degree-of-freedom induced by the brake friction coefficient only.

By considering this system composed of two masses $m_1$ and $m_2$ interconnected by stiffnesses $k_1$ and $k_2$ (Figure 3), the dynamic equilibrium around its static equilibrium position is expressed by the following system of equations

$$\begin{cases} m_1\ddot{X}_1 + c_1(\dot{X}_1 - \dot{X}_2) + k_1(X_1 - X_2) = 0 \\ m_2\ddot{Y}_2 + c_2\dot{Y}_2 + k_2 Y_2 = -N\sin\theta + T\cos\theta \\ m_2\ddot{X}_2 + c_1(\dot{X}_2 - \dot{X}_1) + k_1(X_2 - X_1) = N\cos\theta + T\sin\theta \end{cases} \quad (2)$$

By applying the hypothesis of maintained contact between the mass $m_2$ and the moving belt, the geometric constraint imposes

$$X_2 = Y_2 \tan\theta \quad (3)$$

By eliminating $x$ in equations (3) and considering Coulomb's friction law $T = \mu_f N$, the 2-degrees-of-freedom system has the form

$$\mathbf{M}\ddot{\mathbf{x}} + \mathbf{C}\dot{\mathbf{x}} + \mathbf{K}\mathbf{x} = \mathbf{0} \quad (4)$$

where $\mathbf{x} = \{X_1 \ Y_2\}^T$. $\ddot{\mathbf{x}}$, $\dot{\mathbf{x}}$ and $\mathbf{x}$ are the acceleration, velocity, and displacement response 2-dimensional vectors of the degrees-of-freedom, respectively. The mass matrix $\mathbf{M}$, the damping matrix $\mathbf{C}$ and the stiffness matrix $\mathbf{K}$ of the system are given by

$$\mathbf{M} = \begin{bmatrix} m_1 & 0 \\ 0 & m_2(\tan^2\theta + 1) \end{bmatrix} \quad (5)$$

$$\mathbf{C} = \begin{bmatrix} c_1 & -c_1 \tan\theta \\ c_1(-\tan\theta + \mu_f) & c_1(\tan^2\theta - \mu_f \tan\theta) + c_2(1 + \mu_f \tan\theta) \end{bmatrix} \quad (6)$$

$$\mathbf{K} = \begin{bmatrix} k_1 & -k_1 \tan\theta \\ k_1(-\tan\theta + \mu_f) & k_2(1 + \mu_f \tan\theta) + k_1(\tan^2\theta - \mu_f \tan\theta) \end{bmatrix} \quad (7)$$

Finally, the dynamic system may be rewritten in state variables:

$$\dot{\mathbf{z}} = \mathbf{A}\mathbf{z} \quad (8)$$

where

$$\mathbf{z} = \begin{bmatrix} \mathbf{x} \\ \dot{\mathbf{x}} \end{bmatrix} \quad (9)$$

and

$$\mathbf{A} = \begin{bmatrix} \mathbf{0} & \mathbf{I} \\ -\mathbf{M}^{-1}\mathbf{K} & -\mathbf{M}^{-1}\mathbf{C} \end{bmatrix} \quad (10)$$

## 3.2 CONTROLLED SYSTEM BY APPLYING µ-SYNTHESIS

µ-synthesis is applied to the brake system by assuming that the friction coefficient $\mu_f$ is uncertain. This uncertainty corresponds to possible variations of the friction with time. This controlled system is illustrated in Figure 3(B): the controller $u$ is placed in parallel with the suspension.

### 3.2.1 Definition of the controlled system

By considering section 2 and equations (15-17) of the initial brake system, the nominal system is represented by:

$$\begin{cases} \dot{\mathbf{z}} = \mathbf{A}\mathbf{z} + \mathbf{B}_1\mathbf{u}_1 \\ \mathbf{y}_1 = \mathbf{C}_1\mathbf{z} + \mathbf{D}_{11}\mathbf{u}_1 \end{cases} \quad (11)$$

where

$$\mathbf{B}_1 = \begin{bmatrix} 0 \\ 0 \\ 0 \\ \dfrac{1 + \mu_f \tan\theta}{m_2(\tan^2\theta + 1)} \end{bmatrix} \quad (12)$$

$$\mathbf{C}_1 = \begin{bmatrix} 0 & 1 & 0 & 0 \end{bmatrix} \quad (13)$$

$$\mathbf{D}_{11} = 0 \quad (14)$$

In the presence of uncertainties, the nominal system is modified to introduce the variables corresponding to the uncertain parameters:

$$\begin{cases} \dot{\mathbf{z}} = \mathbf{A}\mathbf{z} + \mathbf{B}_0\mathbf{u}_0 + \mathbf{B}_1\mathbf{u}_1 \\ \mathbf{y}_0 = \mathbf{C}_0\mathbf{z} + \mathbf{D}_{00}\mathbf{u}_0 + \mathbf{D}_{01}\mathbf{u}_1 \\ \mathbf{y}_1 = \mathbf{C}_1\mathbf{z} + \mathbf{D}_{10}\mathbf{u}_0 + \mathbf{D}_{11}\mathbf{u}_1 \end{cases} \quad (15)$$

In the case under consideration, the uncertainty is introduced on the friction coefficient $\mu_f$. As there is only one uncertainty, the set $\Delta_{set}$ is reduced to scalar variables, which elements are noted $\delta$. We have then $\tilde{\mu}_f = \mu_f(1+\delta)$, where $\delta$ is the degree of uncertainty. The matrix $\mathbf{A}$ is then transformed to $\tilde{\mathbf{A}} = \mathbf{A} + \delta.\bar{\mathbf{A}}$, where $\mathbf{A}$ is the previous matrix and $\bar{\mathbf{A}}$ is defined by:

$$\bar{\mathbf{A}} = \begin{bmatrix} 0 & 0 & 0 & 0 \\ 0 & 0 & 0 & 0 \\ 0 & 0 & 0 & 0 \\ -\dfrac{\mu_f k_1}{m_2(\tan^2\theta+1)} & \dfrac{\mu_f \tan\theta(k_1-k_2)}{m_2(\tan^2\theta+1)} & -\dfrac{\mu_f c_1}{m_2(\tan^2\theta+1)} & \dfrac{\mu_f \tan\theta(c_1-c_2)}{m_2(\tan^2\theta+1)} \end{bmatrix} \quad (16)$$

By definition, we have $\mathbf{u}_0 = \delta.\mathbf{y}_0$, which results in: $\delta.\bar{\mathbf{A}} = \mathbf{B}_0.(\delta^{-1} - \mathbf{D}_{00})^{-1}.\mathbf{C}_0$. This relation allows us to determine the matrices $\mathbf{B}_0$, $\mathbf{C}_0$ and $\mathbf{D}_{ij}$:

$$\mathbf{B}_0 = \begin{bmatrix} 0 \\ 0 \\ 0 \\ \dfrac{1}{m_2(\tan^2\theta+1)} \end{bmatrix} \quad (17)$$

$$\mathbf{C}_0 = \begin{bmatrix} -\mu_f k_1 & \mu_f \tan\theta(k_1-k_2) & -\mu_f c_1 & \mu_f \tan\theta(c_1-c_2) \end{bmatrix} \quad (18)$$

$$\mathbf{D}_{00} = 0 \quad (19)$$

$$\mathbf{D}_{01} = 0 \quad (20)$$

$$\mathbf{D}_{10} = 0 \quad (21)$$

The relations between $\mathbf{z}$, $\mathbf{y}_0$, $\mathbf{y}_1$, $\mathbf{u}_0$ and $\mathbf{u}_1$ determined in this section are the basis of the μ-synthesis resolution.

### 3.2.2 Resolution

As explained previously, equations (16-21) correspond to the complete description of the controlled system and contain the nominal system and the perturbations linked to the uncertainties from a general standpoint.

The D-K iterations, allowing us to determine the robust controller by μ-synthesis, are conducted using Matlab software [28]. The controller is determined where the structured singular value $\mu$ of the system is less than unity and we obtain the Bode diagram of the controller allowing robust stability of the system.

The structured singular value $\mu$ obtained for the brake system is plotted in Figure 4. We observe that $\mu$ is less than unity for all frequencies $\omega$, which means that robust stability is assured. This result is obtained after two D-K iterations.

The controller determined by the algorithm and corresponding to this result is illustrated in Figure 5. The controller is approximated by a function $G(\omega)$ which is sought in the form:

$$G(\omega) = -\alpha_G + \beta_G \omega^2 \quad (22)$$

where $\alpha_G$ and $\beta_G$ are constants, depending on the values of the parameters of the system.

For the configuration $\omega_1 = 387$ rad s$^{-1}$, $\omega_2 = 316.2$ rad s$^{-1}$, $\zeta_1 = 0.008$, $\zeta_2 = 0.0065$ (i.e. $m_1 = 1$ kg, $m_2 = 1$ kg, $c_1 = 5$ N m$^{-1}$ s$^{-1}$, $c_2 = 5$ N m$^{-1}$ s$^{-1}$, $k_1 = 1.5 \times 10^5$ N m$^{-1}$, $k_2 = 1 \times 10^5$ N m$^{-1}$), $\theta = 0.2$ rad, $\mu_f = 0.3$, the numerical results give $\alpha_G = 1 \times 10^5$ and $\beta_G = 0.9395$. This approximation of $G(\omega)$ allows a good representation of the controller, as we can see in Figure 6, and it will be useful in the stability analysis of the controlled system.

### 3.3 STABILITY ANALYSIS OF THE INITIAL AND CONTROLLED SYSTEMS

In this section, the stability of the initial and controlled brake systems will be compared. To examine the stability of the initial system, the eigenvalues λ of the matrix **A** (defined in equation (17)) need to be determined. As long as the real part of all the eigenvalues λ remains negative, the system is stable. When at least one of the eigenvalues has a positive real part, the system is unstable. Moreover, the imaginary part of the eigenvalue having a positive real part represents the frequency of the unstable mode.

For the controlled system, the form $G(\omega) = \alpha_G + \beta_G \omega^2$ means that $\alpha_G$ is equivalent to a stiffness and $\beta_G$ is equivalent to a mass. It enables us to take this into account directly in the mass and stiffness matrices. By noting $\tilde{\mathbf{K}}$ and $\tilde{\mathbf{M}}$ the new mass and stiffness matrices

$$\tilde{\mathbf{K}} = \mathbf{K} + \begin{bmatrix} 0 & 0 \\ 0 & 1 + \mu_f \tan\theta.\alpha_G \end{bmatrix} \quad (23)$$

$$\tilde{\mathbf{M}} = \mathbf{M} + \begin{bmatrix} 0 & 0 \\ 0 & 1 + \mu_f \tan\theta.\beta_G \end{bmatrix} \quad (24)$$

a new matrix $\tilde{\mathbf{A}}$ may be defined for the controlled system by

$$\tilde{\mathbf{A}} = -\begin{bmatrix} \mathbf{0} & \mathbf{I} \\ -\tilde{\mathbf{M}}^{-1}\tilde{\mathbf{K}} & -\tilde{\mathbf{M}}^{-1}\mathbf{C} \end{bmatrix} \quad (25)$$

The advantage of these notations is that the stability analysis is similar for the initial and the controlled systems: the sign of the real part of the eigenvalues $\tilde{\lambda}$ of $\tilde{\mathbf{A}}$ gives the result concerning the stability of the system.

First, the evolutions of frequencies in relation to the brake friction coefficient for the initial and controlled brake systems are given in Figure 7. The evolutions of the associated real parts and the representation in the complex plan are given in Figures 8-9. As illustrated in Figure 8, Hopf bifurcation points occur at $\mu_{f0} = 0.35$ and $\tilde{\mu}_{f0} = 0.46$ for the initial and controlled systems respectively. A Hopf bifurcation point is defined by the following conditions

$$\begin{cases} \operatorname{Re}(\lambda_{center}(\mu))\big|_{\mu_f = \mu_{f0}} = 0 \\ \operatorname{Re}(\lambda_{non-center}(\mu))\big|_{\mu_f = \mu_{f0}} \neq 0 \\ \dfrac{d}{d\mu}(\operatorname{Re}(\lambda(\mu)))\bigg|_{\mu_f = \mu_{f0}} \neq 0 \end{cases} \quad (26)$$

where $\lambda_{center}$ defines a pair of purely imaginary eigenvalues while all of the other eigenvalues $\lambda_{non-center}$ have nonzero real parts at $\mu_f = \mu_{f0}$. The last condition of equation (34), called a transversal condition, implies a transversal or nonzero speed crossing of the imaginary axis.

If $\mu_f < \mu_{f0}$ the initial system is stable; it has two stable modes at different frequencies, as illustrated in Figure 7. As the brake friction coefficient increases, these two modes move closer until they reach the bifurcation zone. We obtain the coalescence for $\mu_f = \mu_{f0}$ of two imaginary parts of the eigenvalues. Finally, the initial system becomes unstable for $\mu_f > \mu_{f0}$. In the case of the controlled brake system, the stable and unstable regions are obtained for $\mu_f < \tilde{\mu}_{f0}$ and $\mu_f > \tilde{\mu}_{f0}$, respectively.

In Figure 8, it may be observed that the instability region versus the friction coefficient $\mu_f$ is smaller for the controlled system than for the initial system ($\tilde{\mu}_{f0} > \mu_{f0}$). This illustrates the suitability of robust control via μ-synthesis.

An interesting observation is that the mode that becomes unstable and reaches the bifurcation zone is different for the initial and controlled systems (Figure 11).

Then, in order to demonstrate the suitability of µ-synthesis and in order to compare the stability analysis of the initial and controlled brake systems, different sets of two combinations of physical parameters $k_1$, $k_2$, $c_1$, $c_2$ and $\theta$ are tested. Figures 10-14 show the zones of instability for the initial and controlled systems: the dashed line corresponds to the initial system and the solid line corresponds to the controlled system. Figures 15-16 show the evolutions of the frequencies and the associated real parts in the complex plane.

For all tested combinations of the different parameters with the brake friction coefficient (Figures 10-14), the controller provided by µ-synthesis allows an increased stable zone for the brake system. However, for some values of these parameters which correspond more or less to the nominal setting, this improvement is very weak. Another interesting result of µ-synthesis is that the intervals of instability frequencies are reduced for the controlled brake system in comparison with the initial system, as illustrated in Figures 15-16 and Table 1.

Finally, dynamical responses of the system are presented in Figures 17-18 in order to illustrate the advantages of the controlled brake system versus the initial brake system. In this case, we consider a combination of parameters corresponding to an unstable zone for the initial system and a stable zone for the controlled system ($m_1 = 1 \text{ kg}$, $m_2 = 1 \text{ kg}$, $c_1 = 5 \text{ N m}^{-1} \text{ s}^{-1}$, $c_2 = 5 \text{ N m}^{-1} \text{ s}^{-1}$, $k_1 = 1.5 \times 10^5 \text{ N m}^{-1}$, $k_2 = 1 \times 10^5 \text{ N m}^{-1}$, $\theta = 0.2 \text{ rad}$, $\mu_f = 0.4$). In such a case, the temporal response of the initial system grows exponentially while that of the controlled system is softened (Fig. 17-18). Figure 19 illustrates the oscillations of the initial and controlled systems. It illustrates the difference between the behaviour of the two systems. The instability is manifested by an exponentially increasing curve. On the other hand, the response of the controlled system is more complex but is limited in amplitude.

## 4 CONCLUSION

This study presents an application of µ-synthesis in order to eliminate friction-induced vibration for a brake system. A model for judder instability analysis and an associated stability analysis for the initial and controlled systems are developed. For further understanding of the effects caused by variations in some parameters and the suitability of µ-synthesis, a stability analysis using two parameter evolutions has been conducted.

Robust stability via µ-synthesis for brake systems appears interesting for reducing the size of the instability regions in relation to the possible system parameter combinations. This procedure can be applied to a brake design avoiding friction-induced judder instability.

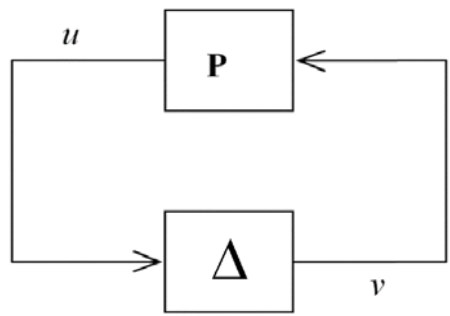

Figure 1 : $\mathbf{P} - \Delta$ feedback connection

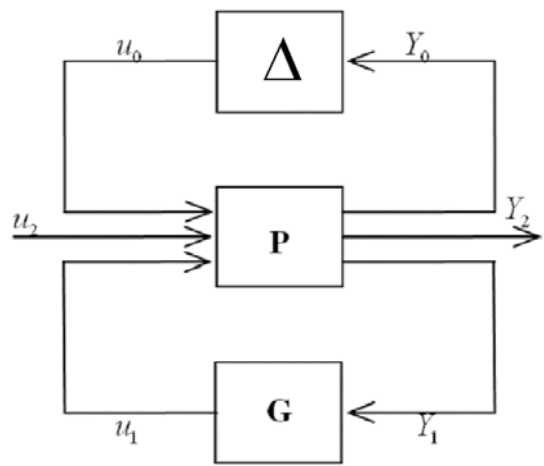

Figure 2 : General structure of the problem

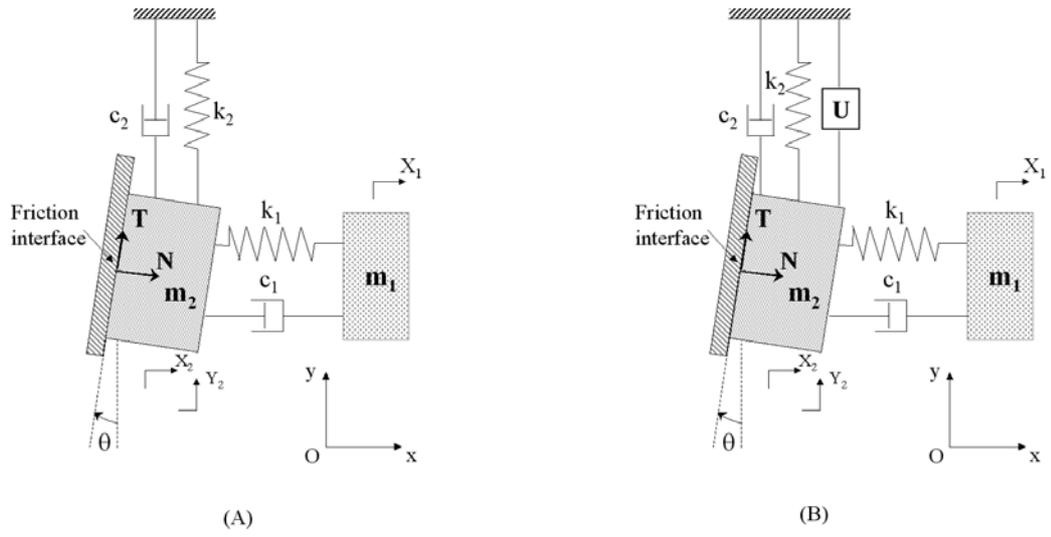

Figure 3 : Braking model (A) initial system (B) controlled system

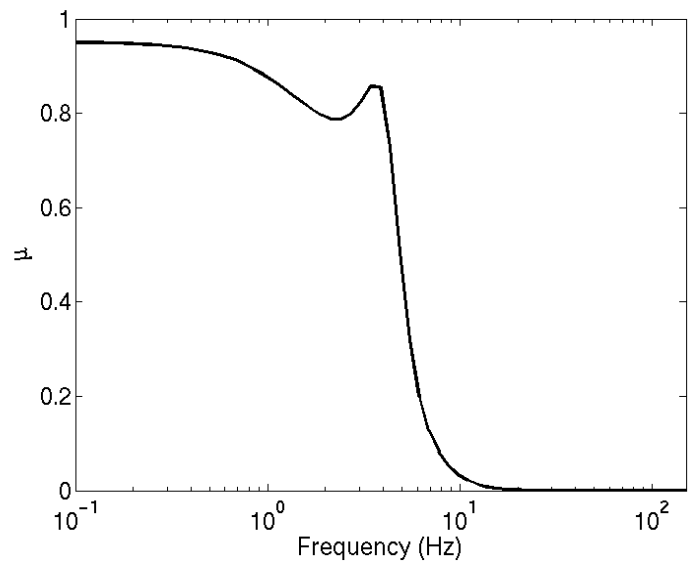

Figure 4 : Evolution of the structured singular value $\mu$

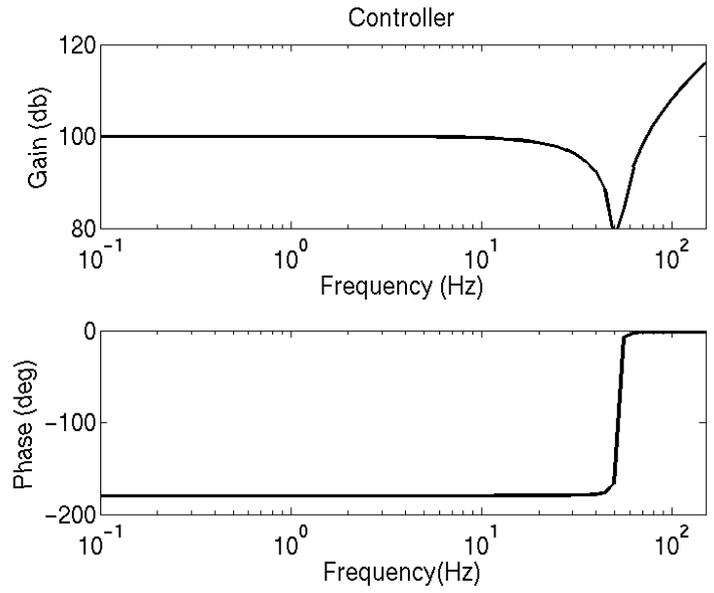

Figure 5 : Bode diagram of the controller

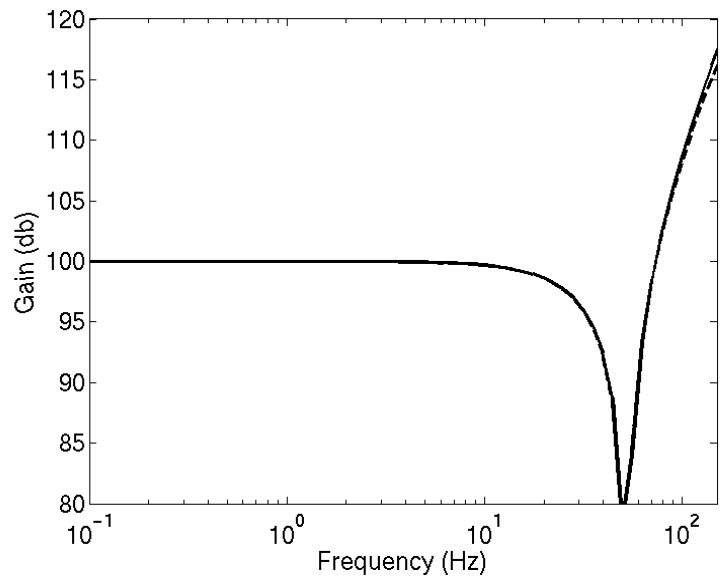

Figure 6 : Approximation of the controller
( ------ Result of the $\mu$-synthesis, ——— Approximation)

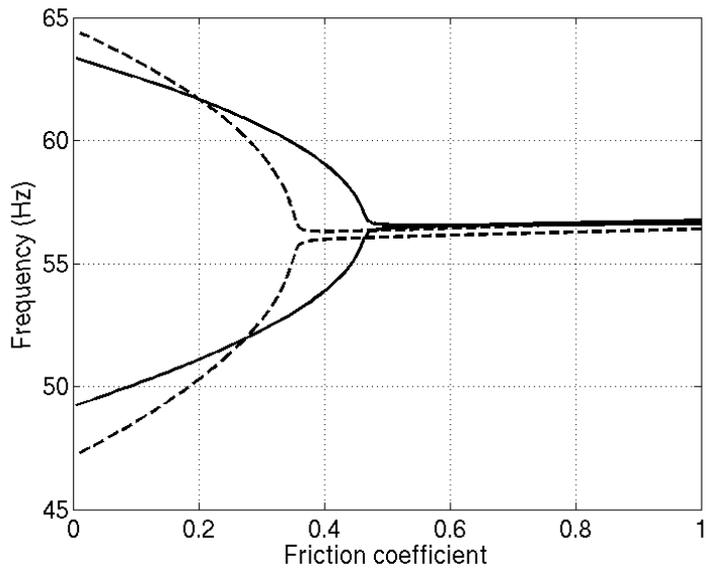

Figure 7 : Evolution of the frequency of two coupling modes
( ------ Initial system, ──── Controlled system)

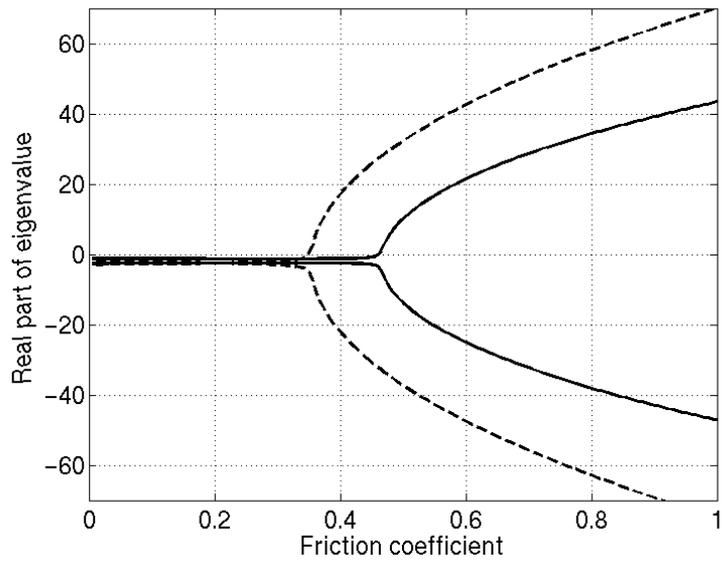

Figure 8 : Evolution of the real part of two coupling modes
( ------ Initial system, ──── Controlled system)

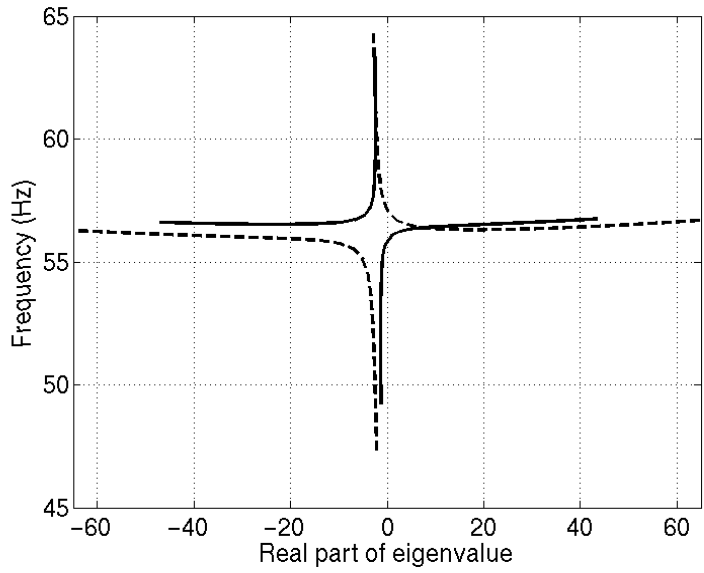

Figure 9 : Evolution of the frequency versus real part of two coupling modes
( ------ Initial system, ──── Controlled system)

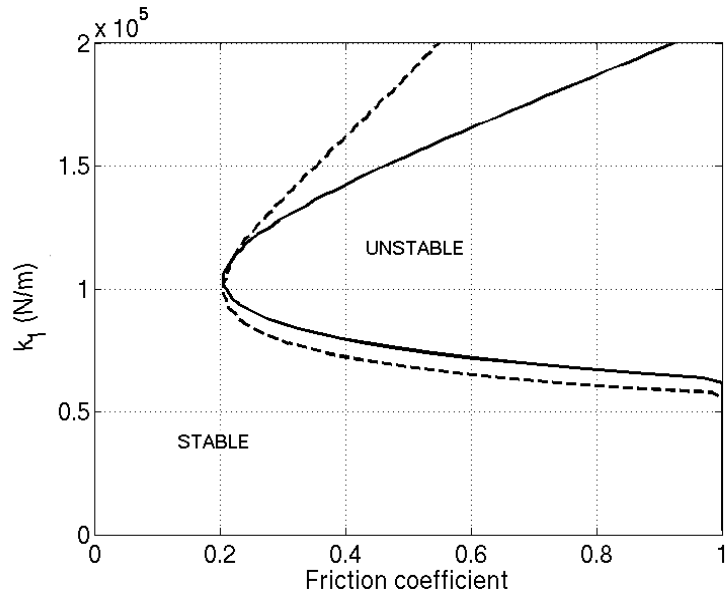

Figure 10 : Stability as a function of brake friction coefficient $\mu_f$ and stiffness $k_1$
( ------ Initial system, ──── Controlled system)

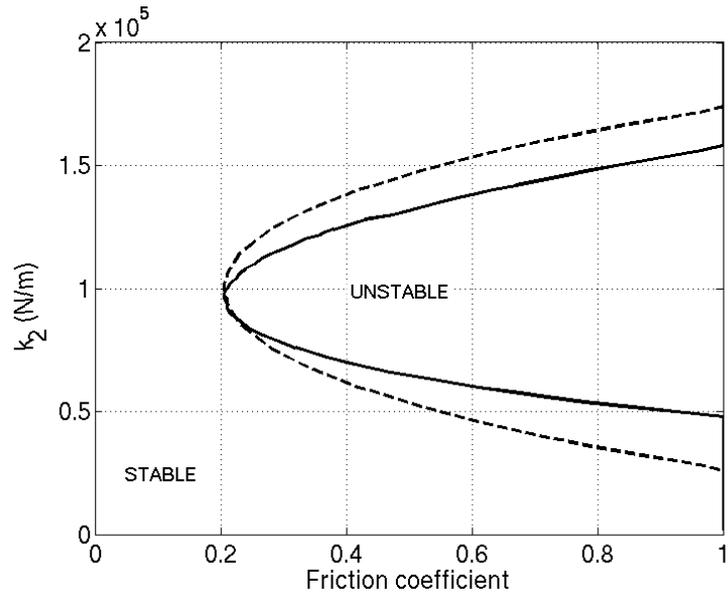

Figure 11 : Stability as a function of brake friction coefficient $\mu_f$ and stiffness $k_2$
( ------ Initial system, ──── Controlled system)

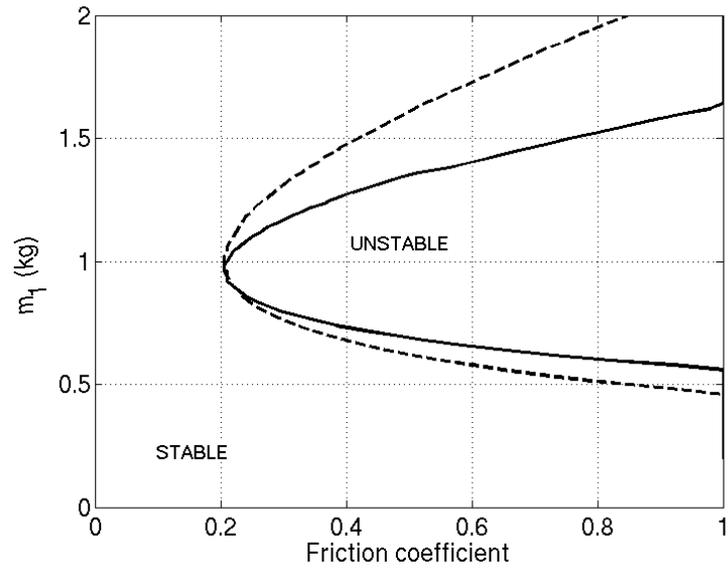

Figure 12 : Stability as a function of brake friction coefficient $\mu_f$ and mass $m_1$
( ------ Initial system, ──── Controlled system)

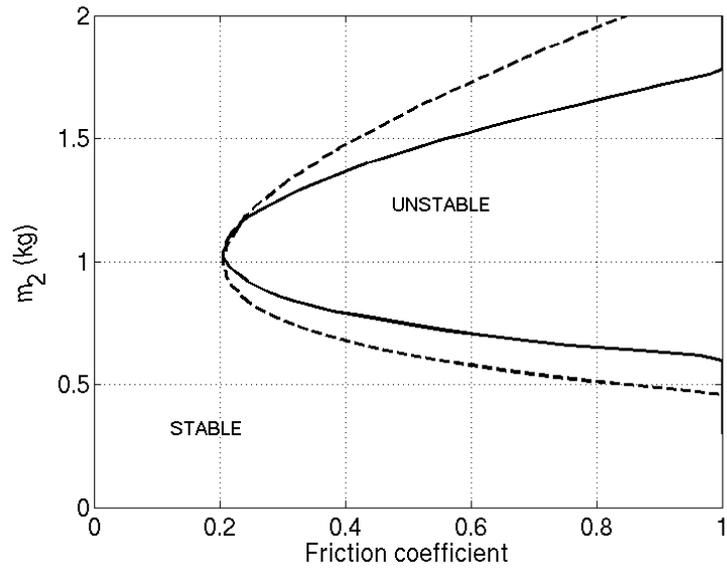

Figure 13 : Stability as a function of brake friction coefficient $\mu_f$ and mass $m_2$
( ------ Initial system, ⎯⎯ Controlled system)

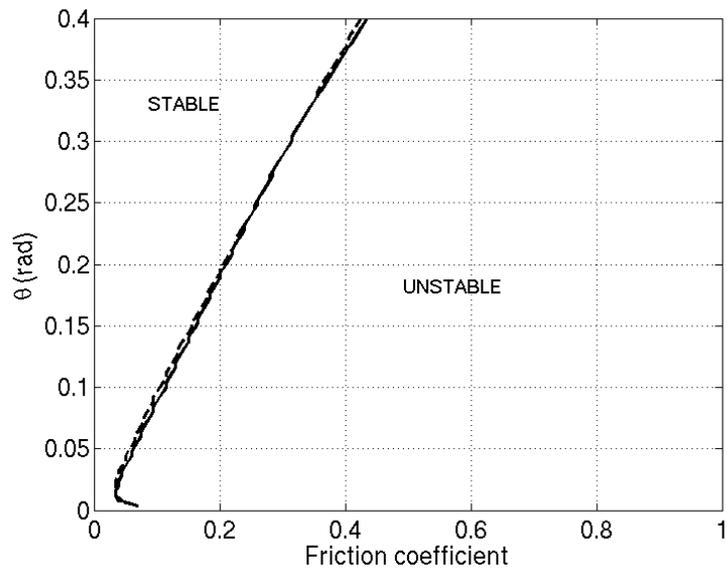

Figure 14 : Stability as a function of brake friction coefficient $\mu_f$ and angle $\theta$
( ------ Initial system, ⎯⎯ Controlled system)

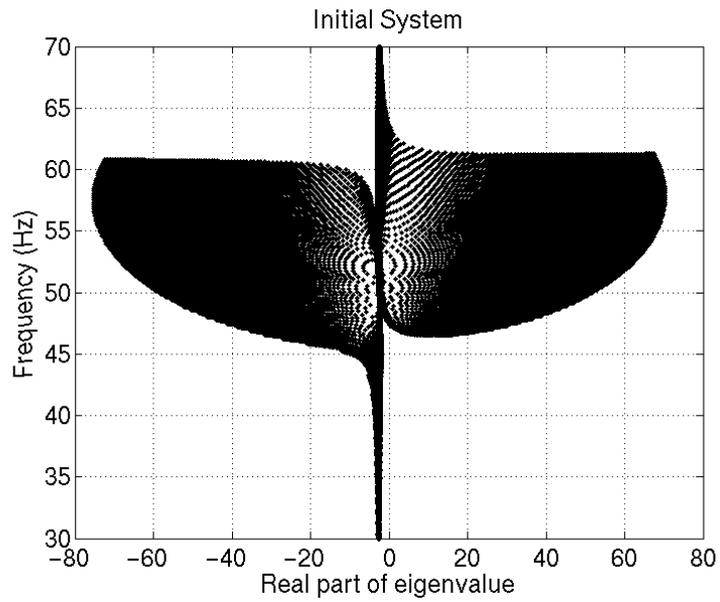

Figure 15 : Frequency and real part of eigenvalue of the initial system for various friction coefficient $\mu_f$ and stiffness $k_1$

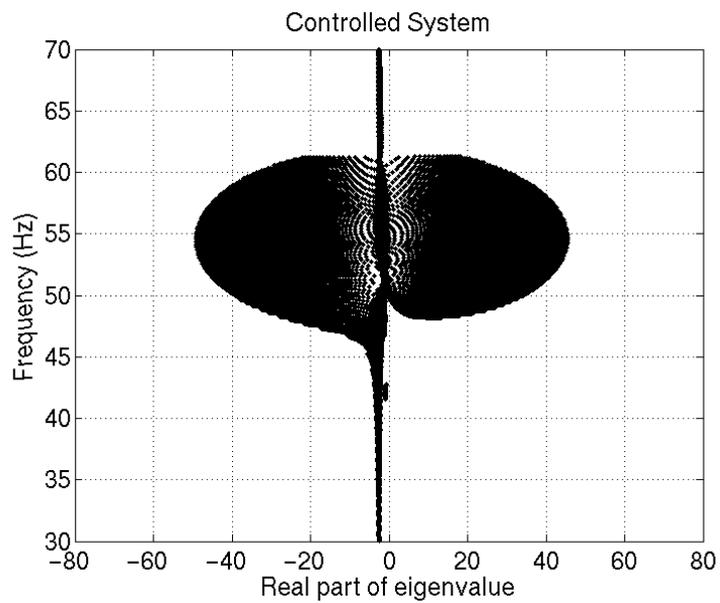

Figure 16 : Frequency and real part of eigenvalue of the controlled system for various friction coefficient $\mu_f$ and stiffness $k_1$

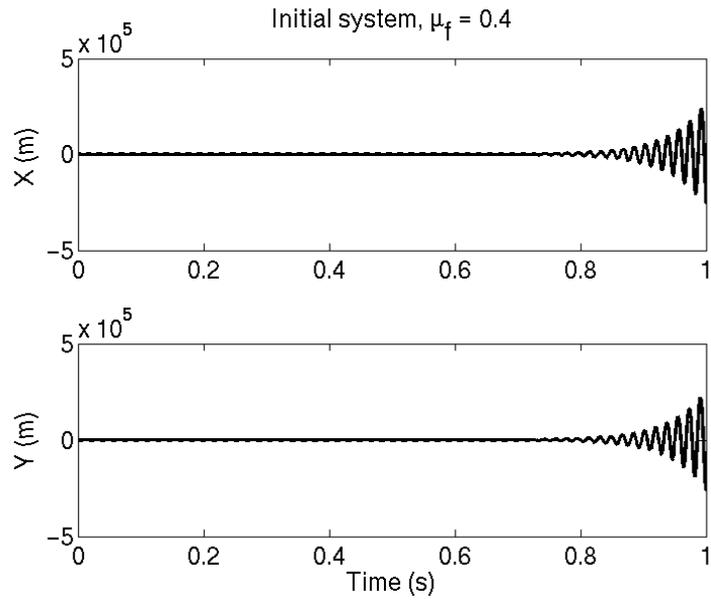

Figure 17 : Temporal response of the initial system for $\mu_f = 0.4$

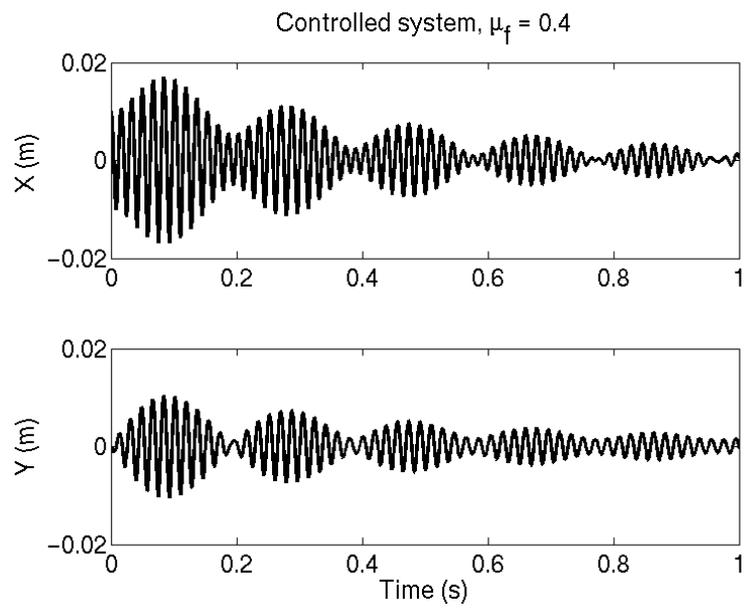

Figure 18 : Temporal response of the controlled system for $\mu_f = 0.4$

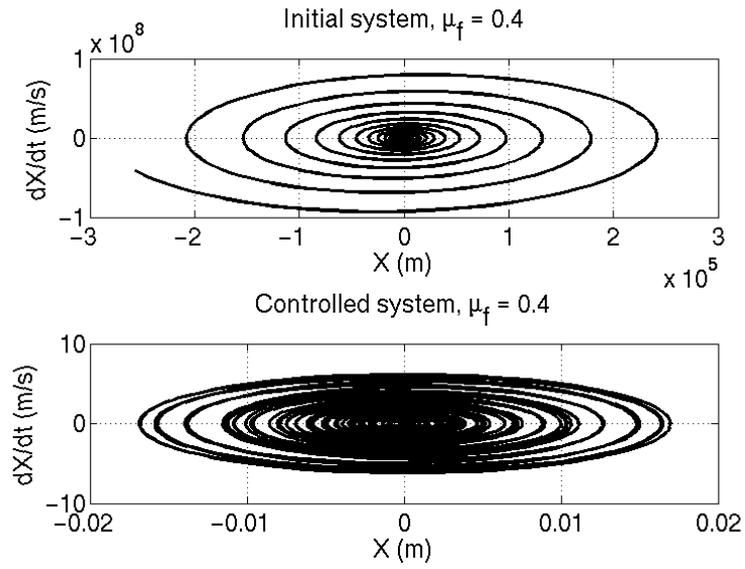

Figure 19 : Divergent oscillations for $\mu_f = 0.4$ $\left(\mu_{f0} < \mu_f < \tilde{\mu}_{f0}\right)$

| Parameter | Initial System | Controlled System |
|-----------|----------------|-------------------|
| $k_1$ | 47-64 Hz | 49-62 Hz |
| $k_2$ | 42-65 Hz | 43-65 Hz |
| $m_1$ | 44-63 Hz | 47-58 Hz |
| $m_2$ | 44-64 Hz | 44-64 Hz |
| $\theta$ | 51 Hz | 49-52 Hz |

Table 1: Comparison of the instability regions for the initial and controlled brake system